\documentclass[graybox]{svmult}
\usepackage{mathptmx}       
\usepackage{bm}
\usepackage{helvet}         
\usepackage{courier}        
\usepackage{type1cm}        
\usepackage{type1cm}    
\usepackage{makeidx}         
\usepackage{graphicx}        
\usepackage{multicol}        
\usepackage[bottom]{footmisc}

\makeindex             
\usepackage{amsmath}

\begin{document}

\title*{Inference of Hubble constant using standard sirens and reconstructed matter density field}
 \titlerunning{Inference of Hubble constant using standard sirens} 
\author{ Supranta S. Boruah \orcidID{000-0003-0985-1488}\\ Ghazal Geshnizjani\orcidID{0000-0002-2169-0579} and \\ Guilhem Lavaux \orcidID{0000-0003-0143-8891} }
\institute{Supranta S. Boruah \at University of Pennsylvania, Philadelphia, PA, United States \email{supranta@sas.upenn.edu}\and Ghazal Geshnizjani \at Perimeter Institute for Theoretical Physics and University of Waterloo,  Waterloo, ON, Canada \email{ggeshnizjani@pitp.ca}
\and Guilhem Lavaux \at Institut d’Astrophysique de Paris (IAP)/CNRS, 98 bis Boulevard Arago, F-75014 Paris, France \email{guilhem.lavaux@iap.fr}}

\maketitle
\abstract*{We summarise a new approach for measuring the Hubble constant using standard sirens and the reconstructed matter density field obtained from observed galaxy surveys. Specifically, we describe and test this method using the Bayesian forward-modelled software \textsc{borg} \cite{borg_original, borg_pm}. This software evolves the initial density field to the present, sampling plausible density fields from the posterior, accounting for peculiar velocities \cite{Hui2006}, and automatically incorporating higher-order correlation functions. The advantage of adding additional information from correlations is expected to make this method more effective in low-signal-to-noise regimes, such as those with modest galaxy number density or incomplete surveys. Our results show that developing a cross-correlation framework between gravitational waves and galaxy surveys, based on the forward-modelled reconstructed density field, to measure $H_0$ is promising. \\ This chapter is based on the research conducted as part of Boruah's Ph.D. thesis \cite{suprantaphd}.}

\abstract{We summarise a new approach for measuring the Hubble constant using standard sirens and the reconstructed matter density field obtained from observed galaxy surveys. Specifically, we describe and test this method using Bayesian forward-modelled software \textsc{borg} \cite{borg_original, borg_pm}. This software evolves the initial density field to the present, sampling plausible density fields from the posterior, accounting for peculiar velocities \cite{Hui2006}, and automatically incorporating higher-order correlation functions. The advantage of adding additional information from correlations is expected to make this method more effective in low-signal-to-noise regimes, such as those with modest galaxy number density or incomplete surveys. Our results show that developing a cross-correlation framework between gravitational waves and galaxy surveys, based on the forward-modelled reconstructed density field, to measure $H_0$ is promising. \\ This chapter is based on the research conducted as part of Boruah's Ph.D. thesis \cite{suprantaphd}.}

\section{Introduction}
\label{intro}
The groundbreaking observation of gravitational waves (GWs) \cite{Abbott:2016blz} resulting from the mergers of stellar binary compact objects ranks among the most extraordinary milestones in contemporary physics. These detections introduce a novel method for exploring space that does not rely on electromagnetic waves or particle emissions. This new avenue offers many opportunities for studying various areas of physics. Of special interest to cosmologists, it presents a novel tool for determining the luminosity distance at cosmological scales, termed \textit{standard sirens} \cite{Holz:2005df, Dalal:2006qt}. This method is especially intriguing because of the increasing conflict between Hubble constant values derived from early and late universe observations. 

Unlike traditional distance measurements from standard candles \cite{Perlmutter:1998np, Riess:2019cxk}, standard sirens do not require intermediate calibration with a distance ladder. Instead, distances are directly estimated using general relativity to calculate a direct distance measurement. The concept of using standard siren to measure the Hubble constant was first proposed by Schutz \cite{Schutz:1986gp}, and has since been refined. 

The first practical application of GW-based cosmological inference came with the detection of a binary neutron star merger event by LIGO, GW170817 \cite{Abbott:2017xzu}, which included an electromagnetic counterpart that helped measure $H_0=70_{-8}^{+12}$ \cite{LIGOScientific:2017zic}. Although this is not yet competitive with values obtained from CMB and SNe observations, it is in agreement with them. This precision could improve significantly if more events with electromagnetic counterparts, such as gamma-ray bursts or kilonovae providing redshift measurements, were observed. Yet, most of the events observed so far match the profile of binary black holes (BBHs) lacking electromagnetic counterparts, \textit{dark sirens}, making it challenging to determine the redshifts of their host galaxies. In fact, as data collection from the LIGO-Virgo-KAGRA network continues to expand and with the forthcoming addition of LIGO-India, a substantial increase in BBH events is anticipated.

However, deficiencies in electromagnetic data could potentially be compensated for by robust statistical methods, provided that effective statistical frameworks are established. For example, one method to start with is to assume that binary events are associated with galaxies in an observational catalogue, and then adjust their redshifts statistically to those of the host galaxies to estimate the Hubble constant or other cosmological parameters. This approach was first applied to the GW170817 data without assuming an electromagnetic counterpart \cite{Fishbach:2018gjp} and later used for dark sirens \cite{LIGOScientific:2021aug}, leading to the estimate $H_0 = 68_{-6}^{+8}$. This statistical method while promising is limited by the ability to identify potential non-visible host galaxies in the siren's sky localization region. Incorrect modelling of the sky localization of the GW data can lead to biased cosmological parameter measurements. Interestingly, we know that galaxies do not have a random distribution. Therefore, alternative methods have been explored to take advantage of large-scale clustering with dark sirens\cite{Oguri:2016dgk, Mukherjee:2018ebj, Mukherjee:2020hyn, Bera:2020jhx}. These methods use the fact that both galaxies and sirens trace the underlying matter distribution, employing correlation functions to statistically infer cosmological parameters.

In what follows, we discuss a new approach that also relies on the assumption that dark sirens trace the underlying matter distribution. The central feature of this approach is to reconstruct an ensemble of density fields consistent with observed galaxy surveys and then apply a bias model to fit the dark siren population, thereby obtaining posteriors for cosmological parameters, specifically $H_0$. That is, statistically measuring the Hubble constant using standard sirens, based on a reconstructed dark-matter density field from a galaxy catalogue. Here, we present a summary of the method skipping some technical details, which can be found in \cite{suprantaphd}.  This reconstruction is performed with the Bayesian forward-modelled software BORG \cite{borg_original, borg_pm}, which uses a physical model of structure formation to evolve the initial density field to the present day, sampling an ensemble of plausible density fields from the posterior. This method accounts for peculiar velocities, which can introduce systematic uncertainties in redshift estimation at low redshifts, and incorporates higher-order correlation functions induced by structure formation.
The forward-modelled reconstruction framework is particularly effective in low signal-to-noise regimes, such as when the number density of the galaxy is modest or surveys are incomplete. Joint reconstruction of the density and true positions of galaxies from photometric surveys has also proven powerful for improving photometric redshift estimates. Therefore, a framework relying solely on the reconstructed density field to measure $H_0$ is promising.

\section{$H_0$ inference from gravitational wave event and galaxy catalogues within BORG Framework}
\label{likelihood+borg}
As mentioned before the goal of this method is to build a probabilistic framework for estimating the Hubble constant ($H_0$) by cross-correlating gravitational wave data with galaxy surveys through a reconstructed density field. To be more specific, use a Bayesian forward modelling software named BORG \cite{borg_original, borg_pm} to reconstruct density fields that optimise the likelihood fit for $H_0$ with given galaxy catalogue data $\{ x_{\text{gal}} \}$ and a set of gravitational wave events $\{x_{\text{GW}}\}$. In the following sections, we describe testing this approach using mock catalogues and actual observed data. However, the first step is to derive a posterior probability $P(H_0 | \{x_{\text{GW}}\}, \{x_{\text{gal}}\})$ or alternatively a likelihood $P( \{x_{\text{GW}}\}| H_0 , \{x_{\text{gal}}\})$ using the Bayes rule, which would be compatible with the reconstructed matter density field, $\delta$, form BORG. 

We start by assuming that the likelihoods of individual standard siren events are independent of each other, and the final joint likelihood can be expressed as a product of individual event likelihoods. The data in the galaxy catalogues from observation provide their redshifts and angular positions of the galaxies, whereas the GW events provide luminosity distances from merger events, and the relation between these coordinates depends on $H_0$. Moreover, the likelihood of dark sirens is influenced by the clustering details (density reconstruction) derived from the galaxy catalogue data. Therefore, the posterior can be written as 
\begin{align}
    P(H_0 | \{x_{\text{GW}}\}, \{x_{\text{gal}}\}) \propto P(H_0) P(\{x_{\text{GW}}\}| \{x_{\text{gal}}\} , H_0) \nonumber \\
    = P(H_0) \prod_{i = 1}^{N_{\text{GW}}} P(x^i_{\text{GW}}| \{x_{\text{gal}}\}, H_0)\,.
\end{align}

Here, $P(H_0)$ is the prior for $H_0$. Now for a specific $x_{\text{GW}}$ event given a specific $\delta$ realization, the probability that the host galaxy is located at comoving coordinate $r$ and angular position $\Omega$ can be written as  
\begin{equation}
\label{eqn:prelike}
     P(x_{\text{GW}}| \{x_{\text{gal}}\}, H_0)\propto P(x_{\text{GW}}|\hat{d}_L(r,H_0), \Omega)  P(r,\Omega|\delta)  P(\delta|\{x_{\text{gal}}\}).
\end{equation}
The probability distribution $P(\delta|\{x_{\text{gal}}\})$, for the observed galaxy catalogue $\{x_{\text{gal}}\}$ is derived from the reconstruction framework within {\sc borg}. {\sc borg} produces a set of density fields $\{\delta\}$ that statistically agree with the galaxy catalogue. 

The probability $P(r, \Omega|\delta)$ in \eqref{eqn:prelike} represents the probability that the host galaxy of the GW event is at the coordinate position $(r, \Omega)$, given a certain over-density profile. This is modeled as 
\begin{equation}
    p(r,\Omega|n_{\text{GW}}, b_{\text{GW}}, \delta) \propto n_{\text{GW}}(1 + \delta_{\text{GW}}(r, \Omega))\,, 
\end{equation}
 where the the new quantity $n_{\text{GW}}$ gives the average rate of mergers per unit volume in the Universe, $\delta_{\text{GW}}$ is the overdensity of merger rate events and $b_{\text{GW}}$, denotes the `bias' parameter quantifying dependence of the merger events overdensity, $\delta_{\text{GW}}$, with respect to the matter overdensity, $\delta$. For our mock simulations, assuming that the binary events are uniformly placed on the galaxies we take $b_{\text{GW}}$ to be the same as galaxy bias and marginalise over that. However, the `bias' of binary events could be a more complicated function of astrophysical properties of host galaxies. 

 The density field is calculated in comoving coordinates $r$ and is expressed in units of $h^{-1}$ pc. However, as pointed out, gravitational wave signals from standard sirens can be used to estimate the luminosity distance to the GW event.
The conversion from $r$ to the luminosity distance, $\hat{d}_L$, depends on $H_0$ through $\hat{d}_L(r,H_0)=r(1+z)/h$ and $z(r,H_0)=H_0 r/c$.


\par To evaluate $P(x_{\text{GW}}|\hat{d}_L(r, H_0), \Omega)$, we implemented the gravitational wave data analysis tool {\sc BAYESTAR} \cite{bayestar}. This application provides a 3D reconstruction of the posterior $P(d_L, \Omega|x_{\text{GW}})$ with respect to $d_L$ and $\Omega$ for given gravitational wave event data. The likelihood can then be obtained using Bayes' theorem as $p(x_{\text{GW}}|d_L, \Omega) \propto \frac{p(d_L, \Omega|x_{\text{GW}})}{p(d_L, \Omega)}$. The prior $p(d_L, \Omega)$ is considered to be a volumetric prior on the luminosity distance, $P(d_L, \Omega) \propto d_L^2$.
 
 Integrating \eqref{eqn:prelike} over the full ensemble of realizations $\{\delta\}$
 the likelihood of a single gravitational wave event can be written as follows:
\begin{align}
   \label{eqn:event_lkl}
    P(x_{\text{GW}}| \{x_{\text{gal}}\}, H_0) \propto \int & d\{\delta\}\, d \Omega\,dr \,P(x_{\text{GW}}|\hat{d}_L(r, H_0), \Omega)  P(r,\Omega|\{\delta\}) \nonumber \\ 
    & P(\{\delta\}|\{x_{\text{gal}}\})\, .
\end{align}
 Lastly, substituting these relations and performing the marginalization over $\delta$ using the Monte Carlo samples supplied by BORG, equation \eqref{eqn:event_lkl}, would lead to
\begin{align}\label{eqn:likelihood2}
     P(x_{\text{GW}}| \{x_{\text{gal}}\} , H_0)  = 
     \frac{H^2_0}{\alpha(H_0)} \int d b_{\text{GW}}d rd \Omega &\Big[ ~ \frac{1}{(1 + z(r, H_0))^2}(1 + \langle \delta^{i}_\text{GW}\rangle )   \nonumber \\ &  P(\hat{d}_L(r,H_0),\Omega|x_{\text{GW}}) P(b_{\text{GW}})\Big]\,,  
     \end{align}
where $\alpha(H_0)$ is the normalization term and $\langle \delta^{i}_{GW}\rangle$ indicates the mean density field from {\sc borg} sampling. We assume that the galaxy catalogue has a high degree of completeness and normalization is only due to the volume effects and $\alpha(H_0) \propto H^3_0$. However, the normalization factor can generally be different corrections to be considered \cite{gw_lkl_normalization, ligo_H0}.
This is the likelihood that was used in the inference calculations. 

\section{BORG methodology}
\label{borg}
We now present an overview of the BORG framework and highlight key information on how the BORG runs were utilised in our study. For a more comprehensive understanding of BORG, we refer the readers to the original publications \cite{borg_original, borg_pm} and for details of the implementations in this study to \cite{suprantaphd}. 
BORG is an advanced software designed to simultaneously achieve multiple objectives with respect to an observational galaxy catalogue. In particular, for our purposes, it can generate a posterior distribution for the large-scale density field of the universe $\{\delta_i\}$, estimate the parameters of the bias model $\Theta_{\textrm{bias}}$, and apply the corrections due to peculiar velocities along the way.

The approach implemented in BORG is forward modelling, which, as the name indicates, infers the late-time nonlinear density field $\{\delta_i\}$ starting from a sample for the initial density field $\{\delta_i^{\textrm{IC}}\}$ and based on a structure formation model. That is, the initial densities, $\delta_i^{\textrm{IC}}$ are forwarded through the structure formation process in a simulation box to the late-time densities, $\delta_i$. Therefore, $\{\delta_i\}$ is some deterministic function of $\{\delta_i^{\textrm{IC}}\}$. 
The density field profile is constructed in the simulation box so that the density in each voxel is one parameter. 
To efficiently sample such a high-dimensional parameter space, an efficient sampling scheme known as Hamiltonian Monte Carlo (HMC) \cite{hmc1, hmc2} is incorporated. 
  
Associating the number of galaxies/halos to the density field requires a bias model. The bias model implemented in this study is based on the nonlinear model proposed in \cite{Neyrinck_bias}, such that the number of observed galaxies are Poisson samples from the density field $\delta$ with Poisson rate $\lambda_g(\delta)$, which has a power law term times an exponential cutoff in order to capture the low number density in the void regions. 
To ensure consistent sampling of both bias parameters and densities, a block sampling algorithm is used. That is, given the observed galaxy numbers $\{ N^{\textrm{obs}}_i \}$ in each voxel, initially the bias parameters $\Theta_{\textrm{bias}}$ are kept constant while the density field is sampled from the conditional posterior $\mathcal{P}(\{\delta^{\textrm{IC}}_i\}| \{ N^{\textrm{obs}}_i \}, \Theta_{\textrm{bias}})$. Subsequently, using the updated density field $\{\delta^{\textrm{IC}}_{i+1}\}$, the bias parameters are conditionally sampled from the posterior bias $\mathcal{P}(\Theta_{\textrm{bias}}|\{N^{\textrm{obs}}_i\},\{\delta_{i+1}\})$. 

Also, as we pointed out earlier in the process of simulation from the initial position of the particles to the final positions, their velocity is also computed, and given that the observation of galaxies provides their redshift, BORG estimates the density in redshift space and systematically addresses the redshift space distortions in the reconstruction.

\section{Testing the $H_0$ inference model using mock data sets}
\label{mocks}
To test our methodology, we applied it to a simulated data set consisting of a mock galaxy catalogue and mocked gravitational wave data corresponding to a sample of 500 sirens hosted in the galaxy catalogue. In this section, we will start by providing the details of the simulated galaxy catalogue and the BORG run executed on this catalogue to reconstruct the density field. Following that, we will explain our approach for simulating the GW data employed in the inference, and then present the results of using our inference framework on these simulated catalogues.
\subruninhead{ Mock galaxy catalogue}
The mock galaxy catalogue was generated using the \verb|VELMASS| simulation suite \cite{Ramanah:2019cbm}. It consists of 10 cosmological simulations probing variations of cosmological parameters and different initial phases. In this work, we have just used the central simulation results, which assume Planck 2015 \cite{Ade:2015xua} values for the cosmological parameters: $\Omega_m = 0.315$, $\Omega_b = 0.049$, $H_0 = 68 $ km s$^{-1}$ Mpc$^{-1}$, $\sigma_8 = 0.81$, $n_s = 0.97$, and $Y_{\text{He}} = 0.248$. The simulated volume is a cubic box with a side length of $2000~ h^{-1}$ Mpc and contains $2048^3$ dark matter particles. It was initialised at redshift of $z = 50$ and evolved to the present epoch using \verb|GADGET2| \cite{Springel:2005mi}. Then, the \verb |ROCKSTAR | halo finder algorithm \cite{Behroozi} was used to find the halos of the final density profile, resulting in a halo catalogue. Afterwards, a smaller box of size $(500~h^{-1}$ Mpc$)^3$ was extracted from the full simulation volume. Within this smaller box, individual galaxies were assigned to halos using an abundance matching technique \cite{HAM1, HAM2} which involves ranking halos and galaxies by halo mass and galaxy brightness, respectively, in descending order and pairing them accordingly. The galaxy luminosities were derived from a Schechter function, based on the best-fit Schechter parameters reported by \cite{2mpp}. Lastly, galaxies fainter and with apparent magnitude above the threshold of $12.5$ were excluded from the dataset, resulting in the mock data for the ``observed" galaxy catalogue, $\{x_{\text{gal}}\}$, with $\sim 8 \times 10^{4}$ galaxies.

\subruninhead{Mock GW catalogue} To produce the GW catalogue, $\{x_{\text{GW}}\}$, we first randomly select $500$ galaxies within $100~h^{-1}$ Mpc from our mock catalogue as host galaxies for GW events. Assuming the Hubble constant of $H_0 = 68$ km s$^{-1}$ Mpc$^{-1}$, the luminosity distances to these sources were calculated. Given these distances, we then simulated GW data events $\{x_{\text{GW}}\}$ corresponding to $1.4 M_{\odot}$-$1.4 M_{\odot}$ binary systems one in each host galaxy using {\sc bayestar} software \cite{bayestar}. The inclinations for the binaries were randomly selected, and only events with a signal-to-noise ratio greater than $12$ were kept in the ``observed" GW catalogue. Note that {\sc bayestar} software has dual application, allowing it to be used in this step as part of the Mock GW data production, but, as pointed out earlier in Section\eqref{likelihood+borg} and explained below, it can also be used as part of the inference process. 

\subruninhead{$H_0$ inference}
Having generated the mock data for $\{x_{\text{gal}}\}$ and $\{x_{\text{GW}}\}$ we performed the likelihood analysis outlined in Section\eqref{likelihood+borg} on these data set to infer the value of $H_0$ and compared it to injected value of $H_0 = 68$ km s$^{-1}$ in the simulations. 

\begin{figure}
    \centering
    \includegraphics[width=0.7\linewidth]{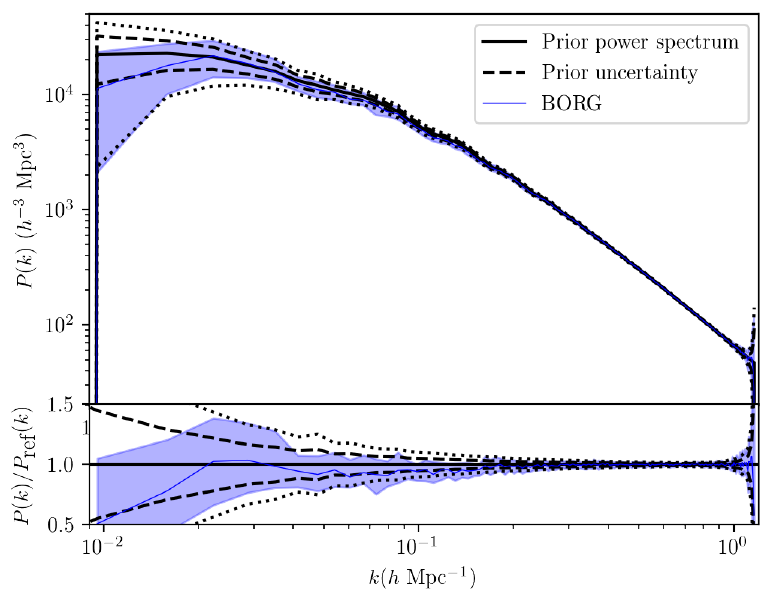}
    \caption{The power spectrum of the initial density fields, adjusted to $z=0$ using linear perturbation theory, is illustrated. The black solid line depicts the $\Lambda$CDM prediction, and the dashed and dotted lines represent the expected $1\sigma$ and $2\sigma$ uncertainties, assuming Poissonian uncertainty while the blue solid line represents the mean of all samples reconstructed by BORG and the blue shaded area indicates the $2\sigma$ area for BORG samples \cite{suprantaphd}.}
    \label{fig:borg_power_spectrum}
\end{figure}

The steps from \eqref{eqn:event_lkl} to \eqref{eqn:likelihood2} involved performing BORG to reconstruct the density field samples corresponding to the mock catalogue $\{x_{\text{gal}}\}$ and marginalizing over them. Given the potential luminosity dependence of the galaxy bias, galaxies with absolute magnitude between $-23.25$ and $-25.25$ were divided into eight equal width bins of absolute magnitude, and the galaxy bias parameters $\Theta_\text{bias}$ for each bin were separately modelled while consistently accounting for red-shift space distortions introduced by peculiar velocity. Figure \ref{fig:borg_power_spectrum} illustrates how the resulting power spectrum from BORG samples compares to the prior $\Lambda$CDM power spectrum used in the $N$-body simulation (scaled to $z=0$ using linear perturbation theory). 

Next, evaluating $P(x_{\text{GW}}|\hat{d}_L(r, H_0), \Omega)$ in \eqref{eqn:likelihood2} as eluded before involved using {\sc bayestar} again but this time as part of the inference process . That is {\sc bayestar} generates a 3 dimensional volume probability for the distance and the sky localization of each of the GW events in the mock $\{x_\text{GW}\}$. This probability distribution was constructed considering two scenarios: {\it i)} Active GW detectors include only the LIGO and VIRGO detectors (HLV), and {\it ii)} in addition to HLV, the LIGO India and KAGRA detectors are also active (HLVIK).

\begin{figure}
    \centering
    \includegraphics[width=\linewidth]{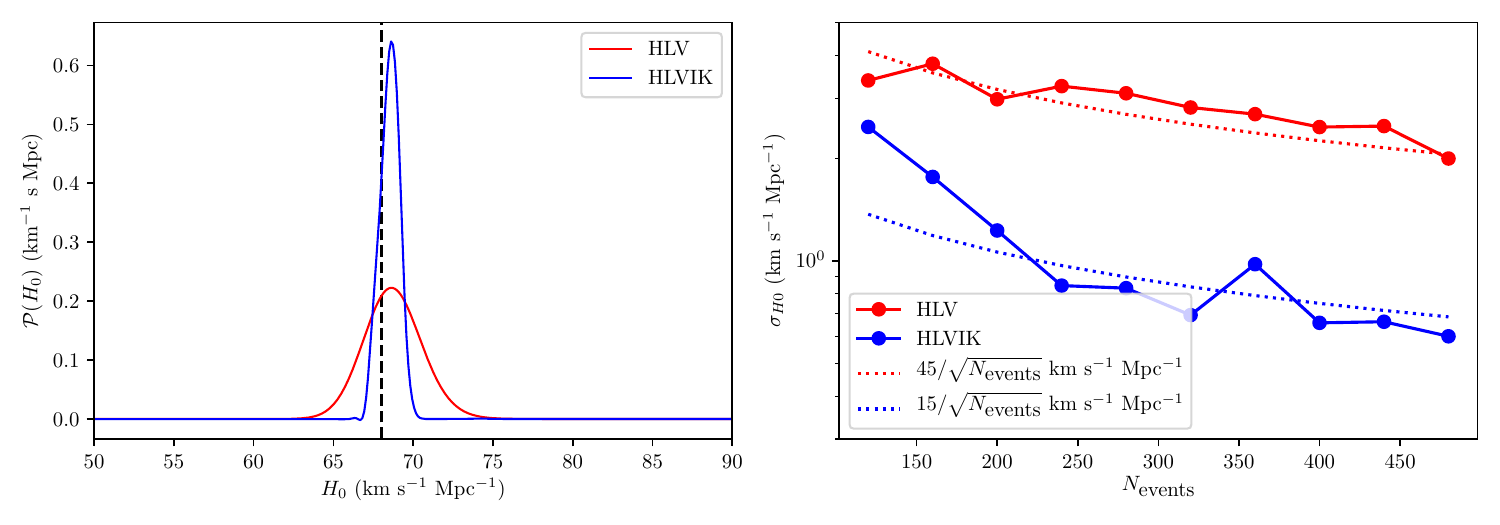}
    \caption{Inference on $H_0$ using the mock $\{x_\text{GW}\}$ and $\{x_\text{GW}\}$ data set. \textit{(Left):} Posterior distribution of $H_0$ for both HLV and HLVIK data sets. The true injected $H_0$ value is marked by the black vertical line. \textit{(Right):} The variation in uncertainty of $H_0$ as a function of the number of events for either cases.}
    \label{fig:H0_mock_inference}
\end{figure}

The posteriors of these data sets for $H_0$ are illustrated in the left panel of Figure \ref{fig:H0_mock_inference}. In both scenarios, an unbiased value of the Hubble constant is recovered. For $500$ HLV events, we obtain $H_0 = 68.75 \pm 1.83$ km s$^{-1}$ Mpc$^{-1}$. Similarly, for $500$ HLVIK events, the recovered value is $H_0 = 68.53 \pm 0.64$ km s$^{-1}$ Mpc$^{-1}$. Moreover, as observed in the right panel of Figure \ref{fig:H0_mock_inference}, with an increase in the number of events, the error appears to decrease approximately in proportion to $1/\sqrt{N_{\text{events}}}$ similar to a Poisson distribution.

\section{$H_0$ inference for GW170817 and 2M++}
\label{GW170817}
Following initial validation with simulated data, we extended our framework's evaluation to real observational data for gravitational waves (GWs) and galaxies. Presently, the computational restrictions of the Bayesian reconstruction from galaxies (BORG) framework impose limitations on the box size, thereby constraining the redshift range. Consequently, our current assessment is restricted to the low-redshift domain. Nevertheless, we anticipate that this method will demonstrate its full potential in the future, given the fast pace of advances in computational efficiency and the availability of extensive high-redshift galaxy surveys and GW data.   
With that in mind, we applied our inference method on galaxy catalogue 2M++ \cite{2mpp} and the GW data from the binary neutron star event GW170817 using BORG reconstruction for the density. 
Although an optical counterpart was identified for GW170817, our analysis does not incorporate this information. An alternative measurement of the Hubble constant, $H_0$, which does not rely on the optical counterpart of GW170817 and correlations to the density field, is described in \cite{Fishbach_hubble}. 

The detailed account of {\sc borg} reconstruction of the 2M++ catalogue is presented in \cite{borg_pm}. The galaxy bias is again modelled assuming luminosity dependence and by splitting galaxies into eight bins of equal width based on their absolute magnitude in this case between $-21$ and $-25$.
Additionally, the galaxies in the 2M++ catalogue with $K_{\text{2M++}} < 11.5$ and those in $11.5 < K_{\text{2M++}} < 12.5$ were categorised separately due to different selections. 

\begin{figure}
    \centering
    \includegraphics[width=0.6\linewidth]{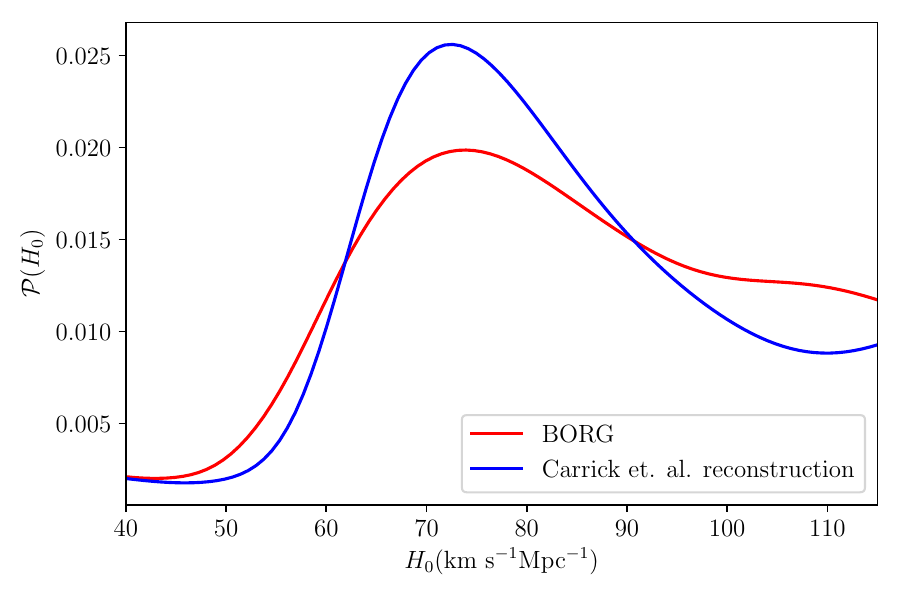}
    \caption{Posterior for $H_0$ given the data corresponding to GW170817 event and 2M++ using two distinct reconstruction techniques. The red curve represents the posterior obtained via the {\sc borg} reconstruction applied to the 2M++ catalogue while the blue curve reconstruction method described in \cite{Carrick_et_al}.}
    \label{fig:gw170817}
\end{figure}

The inference for $H_0$ resulted in $H_0 = 84^{+23}_{-18}$ km s$^{-1}$ Mpc$^{-1}$ (see Figure \eqref{fig:gw170817}), consistent with other cosmological measurement methods for $H_0$, but not offering competitive precision. Furthermore, we also derived a similar result using our approach, but substituting the reconstructed galaxy over-density field based on the luminosity weighting method presented in \cite{Carrick_et_al} in place of the BORG reconstructed matter density. The result was $H_0= 82^{+20}_{-14}$ km s$^{-1}$ Mpc$^{-1}$, which is comparable to BORG (see Figure \eqref{fig:gw170817}) and also aligns with the findings in \cite{Fishbach_hubble}. This is expected, as the matter density correlation function should be more effective at higher redshifts, where survey completeness is more limited.

\section{conclusion}
\label{conclusion}

We presented a new approach for inferring cosmological parameters through the cross-correlation of gravitational wave data with galaxy surveys. More specifically provided an analytical likelihood that can employ a forward-modelled reconstructed density field obtained from observed galaxy surveys via {\sc borg} software and using standard sirens to estimate the Hubble constant ($H_0$). 
This framework automatically incorporates higher-order correlation functions and we predict to be effective in low-signal-to-noise regimes, such as modest galaxy number density or incomplete surveys, where the correlation could provide additional information.  
 We tested this framework on simulated datasets for galaxies and gravitational wave (GW) events, showing that our method provides an unbiased estimate of $H_0$. We also applied it to data from the binary neutron star event GW170817 and density reconstructions from the 2M++ galaxy redshift catalogue using {\sc borg}. Our method consistently predicted $H_0 = 84^{+23}_{-18}$ kms$^{-1}$ Mpc$^{-1}$, in agreement with other cosmological measurements, but not very competitive. 
 We also derived a similar result using our approach, but instead of {\sc borg} substituting the reconstructed galaxy over-density field based on the luminosity weighting method. This strongly suggests that reconstruction methods can provide unbiased inferences. In fact, reconstructing the density field jointly while inferring photometric redshifts is known to improve the photometric redshift estimate \cite{JensPhotoz}. This would integrate well with our method for statistically inferring $H_0$ from high redshift GW events. A similar analysis can be applied to Type Ia supernovae when the host galaxy cannot be uniquely identified. This will be particularly relevant for future large-scale surveys like the Vera-Rubin Observatory (VRO), which will detect a substantial number of supernovae. Future work will need to compare our method with data that will become available in the coming decades for both GWs and galaxy surveys, as well as numerical advancements in high-redshift density reconstruction simulations.

\begin{acknowledgement}
The research carried out during this study by SB and GG was supported by a Discovery Grant from the Natural Science and Engineering Research Council of Canada (NSERC). GG is also partly supported by the Perimeter Institute for Theoretical Physics (PI). Research at PI is supported by the Government of Canada through the Department of Innovation, Science and Economic Development Canada and led by the Province of Ontario through the Ministry of Research, Innovation and Science.
GG is also grateful for invitation to the International Symposium on Relativistic Astrophysics held in Gangtok, Sikkim, India to present this work. 
\end{acknowledgement}
%
\bibliographystyle{unsrt}
\bibliography{biblio}
\end{document}